\documentclass{pasj00}%Produced by Zhao 15/07/2005
\usepackage{graphicx}
\draft
\begin{document}
\SetRunningHead{Zhao, Li \& Wu et al.}{ER UMa: Spectroscopy and
Photometry}
\Received{~~~} \Accepted{~~~}

\title{Superhumps Behavior during Normal Outbursts in ER~ UMa: Spectroscopy and Photometry}

\author{Yinghe Zhao, Zongyun Li, Xiaoan Wu, Qiuhe Peng}
\affil{Department of Astronomy, Nanjing University, Nanjing
210093, China}\email{yhzhao, zyli@nju.edu.cn} \and
\author{Zhousheng Zhang, Zili Li}
\affil{Yunnan Astronomical Observatory, Kunming 650011, China}


\maketitle

\begin{abstract}
We have taken a 4-day spectroscopic observation and have been
conducting a 13-day photometric study of the SU UMa-type dwarf
nova, ER Ursae Majoris. The mean $K$-amplitude for the emission
lines is $54\pm 8$ km s$^{-1}$ and $\gamma$ velocity is 8$\pm 4$
km s$^{-1}$ from our spectroscopic results. A phase shift of 0.22
is also obtained.

Our photometric observation confirms superhumps in normal
outbursts of ER UMa and reconciles contradictory observational
results obtained by different authors. We find that superhumps
possibly develop near each normal outburst maximum and fade out
before the next outburst maximum. If the observed humps were not
late superhumps but ordinary superhumps, much more theoretical
works should be done to explain the new phenomena.
\end{abstract}

\section{Introduction}

Cataclysmic Variables (CVs) are short-period binaries in which the
Roche-Lobe-filling secondary transfers mass through the inner
Lagrangian point. Most CVs are active systems with more or less
change in magnitude at various time scales. SU UMa-type dwarf
novae are defined as a special class of CVs with two distinctive
outbursts, ``normal outbursts" as in SS Cyg stars and
``superoutburst" with larger amplitude and lasting 5-10 times
longer. An excellent review of superoutbursts was presented by
Warner (1995).

It has been extensively accepted that normal outbursts are
triggered by ``thermal instability" described by ``S-curves"
(Meyer \& Meyer-Hofmeister 1981; Warner 1995), which gives the
$\Sigma - T_{eff}$ relationship for each annulus in the accretion
disk. And superoutbursts are explained as the result of
``thermal-tidal instability" (Osaki 1989, TTI hereafter) in the
disk. In this model, the tidal instability is only effective after
the radii of the accretion disk grows larger than a critical
value, which is satisfied exclusively in superoutburst. After the
radius exceeds the critical value, the disk becomes eccentric and
produces superoutbursts and ``superhumps", which are periodic
variations of about 0.2-0.3 mag in light curves after
superoutburst maximum.

Another interesting problem is the explanation of the negative
superhumps. The most plausible model is based on the assumption of
a retrograde precession of a titled accretion disk (Patterson et
al. 1993, 1997). Although the positive superhumps have been
simulated numerically since the work of Whitehurst (1988), there
are some difficulties to simulate the negative sumperhumps. In
fact, all attempts to simulate negative superhumps failed in sense
that they were not able to produce a significant tilt starting
from a disk lying in the orbital plane (Murray \& Armitage 1998;
Wood et al. 2000). Once tilted, however, the accretion disk starts
precession in retrograde direction (Wood et al. 2000). However,
the origin of the disk tilt remains uncertain.

The studies of photometry (Kato \& Kunjaya 1995; Robertson,
Honeycutt, \& Turner 1995) show that ER UMa is a peculiar SU
UMa-type dwarf nova with short supercycles of 40-50 days in which
the superoutburst lasts about 20 days. Its normal outburst has a
period of 4 days. Misselt and Shafter (1995, named paper I
hereafter) detected no superhump in normal outburst (see 1994
March 11 in their Figure 8). Kato and Kunjaya (1995, named paper
II hereafter) also declared that no clear evidence of periodic
humps with an amplitude lager than 0.05 mag was detected during a
normal outburst. However, Gao et al. (1999, named paper III
hereafter) found that superhumps exist in normal outbursts of ER
UMa, which is not expected in TTI model. Although they claimed in
paper III that ``it is more likely that superhumps occasionally
exist at essentially all phases of the eruption cycles of ER UMa
stars", it remains uncertain in the rising to normal maximum
because of lack of observations.

Spectroscopic studies have not been performed as extensively as
photometric ones. Thorstensen et al. (1997, TTBR hereafter)
derived an orbital period of 0.06366 days and a semiamplitude,
$K$, of 48$\pm$4 km s$^{-1}$ of the radial velocity. Both
spectroscopy and photometry performed simultaneously were much
scarcer. We took both kinds of observations in 2004 February, and
in this paper we report the spectroscopic features of the
accretion disk during normal outburst and the superhumps behavior
during the rise to normal outburst maximum and the development in
normal outburst.

\section{Observation}

We took spectroscopic observations of ER UMa for 25.7 hr over 4
nights through February 27 to March 1 in 2004, using the OMR
Cassegrain spectrograph attached to the 2.16 m telescope with a
TEK1024 CCD camera at Xinglong Station of National Astronomical
Observatory of China. A total of 63 useful spectra was obtained.
The technique of observation and data processing is similar to
that in Wu et al. (2001). A 300 groove mm$^{-1}$ grating was used,
and the slit width was set to 2$^{\prime \prime}$. We recorded
dome flats at the begging and end of each night. The lamp spectra
were taken before and after every two star exposures and used to
interpolate the coefficients of wavelength scales. A spectral
resolution of 13 \AA\ was derived from FWHM of the lamp spectra
and the rms of identified lines was 0.19 \AA\ using a fourth-order
Legendre polynomial to fit the lines, corresponding to 12 km
s$^{-1}$ near H$\beta$.

The photometric observation of ER UMa was taken in 1999 and 2004,
using a TEK1024 CCD camera attached to the Cassegrain focus of the
1.0 m reflector at Yunnan Observatory. The observation lasted for
26 hours over 6 nights through December 1999 to January 2000 and
for 40 hours over 7 nights in February 2004, in each night covered
well over 3-4 superhump periods of the object. Exposure times were
40, 60 or 90 s, long enough to assure good signal-to-noise ratio.
A total of 1892 useful object frames were obtained through V
filter. After bias subtraction and flat field correction, we
removed the sky background and measured  magnitudes of ER UMa and
two secondary photometric standards, star 4 and star 10 in finding
chart of Henden \& Honeycutt (1995). We use these two standards as
the comparison star (star 4) and check star (star 10),
respectively, after paper III. The rms error is less than 0.02
mag. The journal of our detailed observations is summarized in
Table 1.

%%%%%%%%%%%%%%%%%%%%%%%Table 1%%%%%%%%%%%%%%%%%%%%%%%%%%%%%%%%%%%%%%%%
\begin{table}
\caption{Journal of observations}
\begin{center}
\begin{tabular}{lcccccc}
\hline \hline
Date&HJD start &Duration &Exposure$^a$ &   & Period &$\Delta$m\\
(UT)&(2,451,000+) & (hr)  & (s) & Plates& (days) &(mag) \\
\hline\noalign{\smallskip}
&&&Spectroscopy\\
\hline
2004 Feb 27 & 2063.0455 & 1.75    &1200   & 7    &   ...         &  ...\\
2004 Feb 28 & 2064.0536 & 7.36    &1200   & 17   &   ...         &  ...\\
2004 Feb 29 & 2065.0425 & 8.40    &900    & 27   &   ...         &  ...\\
2004 Mar 1  & 2066.0344 & 8.21    &1200   & 12   &   ...         &  ...\\
\hline \noalign{\smallskip}
&&&Photometry\\
\hline \noalign{\smallskip}
1999 Dec 30 & 543.178   & 4.56    & 90    & 118  &   0.0659(20)  &  0.12 \\
1999 Dec 31 & 544.182   & 4.35    & 60    & 120  &   0.0656(17)  &  0.30 \\
2000 Jan 1  & 545.176   & 4.97    & 60    & 143  &   0.0643(23)  &  0.47 \\
2000 Jan 2  & 546.177   & 4.51    & 90    & 75   &    ---        &  $\sim$0.03\\
2000 Jan 3  & 547.189   & 4.37    & 90    & 120  &   0.0671(20)  &  0.12 \\
2000 Jan 4  & 548.179   & 5.04    & 90    & 135  &   0.0663(35)  &  0.18 \\
2004 Feb 23 & 2059.396  & 3.83    & 90    &  63  &    ---        &  $\sim$0.15 \\
2004 Feb 24 & 2060.344  & 6.50    & 90,60 & 200  &    ---        &  $\sim$0.1 \\
2004 Feb 25 & 2061.338  & 5.01    & 60,40 & 180  &   0.0686(13)  &  0.05 \\
2004 Feb 26 & 2062.329  & 6.25    & 60    & 209  &    ---        &  $\sim$0.08 \\
2004 Feb 27 & 2063.369  & 6.25    & 90    & 171  &    ---        &  0.1-0.2 \\
2004 Feb 28 & 2064.333  & 6.12    & 90    & 197  &    ---        &  0.1-0.3 \\
2004 Feb 29 & 2065.335  & 6.01    & 90    & 161  &   0.0590(10)  &  0.4-0.15 \\
\hline \noalign{\smallskip}
\end{tabular}
\end{center}
\begin{list}{}{}
\item[$^{\mathrm{a}}$]Exposure time shown here is the most common
one at each night.
\end{list}
\end{table}

\section{Results and analysis}
\subsection{Average spectrum and the radial velocity}
In Figure 1 we show the sum of all 63 individual spectra, which
had been normalized to continuum. No radial velocity shift was
applied during the combination. The spectrum presented strong
Balmer emission lines and weaker He~I lines from $\lambda4471$\AA\
to $\lambda6678$\AA. There is also He~II $\lambda$4686 \AA.
Equivalent widths of spectral features are given in Table 2.

%%%%%%%%%%%%%%%%%%%%Table 2%%%%%%%%%%%%%%%%%%%%%%%%%%%%%%%%%%
\begin{table}
\caption{Equivalent Width of Spectral Lines}
\begin{center}
\begin{tabular}{cccc}
\hline \hline
&Equivalent & &Equivalent\\
Element, & Width &Element,&Width\\
Rest Wavelength& (\AA)  & Rest Wavelength & (\AA) \\
\hline \noalign{\smallskip}
H$\epsilon$ $\lambda$3970 & 16.6   & He I $\lambda$3889  &8.2\\
H$\delta$ $\lambda$4101   & 18.6   & He I $\lambda$4471  &5.1\\
H$\gamma$ $\lambda$4340   & 24.9   & He I $\lambda$4922  &3.7 \\
H$\beta$ $\lambda$4861    & 27.1   & He I $\lambda$5015  &3.8  \\
H$\alpha$ $\lambda$6563   & 33.7   & He I $\lambda$5876  &7.6  \\
He II $\lambda$4686       & 3.9    & He I $\lambda$6678  &4.4 \\
\hline \noalign{\smallskip}
\end{tabular}
\end{center}
\end{table}

We measured the central wavelengths of H$\beta$ emission peaks
with the Gaussian-fit method. Figure 2 shows the velocities folded
on the orbital phase with the best-fit sinusoidal superposed. The
orbital phase was calculated according to the ephemeris given by
TTBR,
\begin{equation}
T_0=HJD2449740.0478(8)+0.06366(3)E
\end{equation}
where $T_0$ is the time of the $\gamma$ crossover from negative to
positive, and $E$ is the cycle number. The best-fit sinusoidal
shows that the H$\beta$ emission has $K$=54$\pm$8 km s$^{-1}$,
$\gamma$=8$\pm$4 km s$^{-1}$. TTBR obtained smaller $K$, 48$\pm4$
km s$^{-1}$ and smaller $\gamma$, -31$\pm$3 km s$^{-1}$ from the
H$\alpha$ emission lines using two-Gaussian convolution method
with a separation of 1260 km s$^{-1}$. The large difference of
$\gamma$ should be mainly caused by the following reason (see next
paragraph). The large phase offset of 0.22 shows that the
variation of line peaks can not exactly represent the motion of
the white dwarf.

The symbols in Figure 2 represent different nights. It is clear
that there exist systematic discrepancies between different
nights. The distribution of velocities on February 27 is almost
below the sinusoidal. Those on February 28 and 29 moved upper and
upper with time, and then down again on March 1. The mean values
of velocities on different nights are -64 km s$^{-1}$, 4 km
s$^{-1}$, 38 km s$^{-1}$ and -10 km s$^{-1}$, following the date
sequence. This result is very reasonable if we accept that the
outer part of the accretion disk is eccentric and precessing. A
detailed analysis of variation of mean velocity produced by an
eccentric precessing outer disk at different phase was given by Wu
et al. (2001) and Zhao et al. (2005, 2006). Since our data were
all taken in a normal outburst (see $\S$3.2.2), it is naturally
believed that the outer disk of ER UMa was eccentric and
precessing during its \emph{normal} outburst. Hence, it is not
surprised that the $\gamma$ velocity of TTBR, which is measured in
quiescence, is much smaller than that of ours.

\subsection{Superhumps in the decline part of outbursts}
\subsubsection{The outbursts occurred in 1999-2000}
Figure 3a shows the photometric data recorded in December 1999 and
January 2000. The light curve covers the decline part of an
outburst and another almost complete outburst. The amplitude of
the second outburst is about 2.2 mag and it lasts about 4 days.
But we could not make sure whether these two outbursts are normal
outburst or superoutburst. For the first outburst, it is
incompletely observed and it may be a superoutburst or a normal
outburst.
%One way used to distinguish the late superhumps from
%ordinary superhumps is to calculate the O-C diagram of the
%modulations (vogt 1983; Hessman et al. 1992). If the phase of
%superhump period shifts 180$^\circ$, it is a late superhump.
%However, it is very hard to calculate the O-C diagram for our
%observational data because of the complicated behavior (for
%example, both positive and negative superhumps were found, Paper
%III).
For the second outburst, the mean decline rates are 0.73 mag
day$^{-1}$ and 0.95 mag day$^{-1}$, for January 3 and 4,
respectively. These values are similar to those of normal
outbursts presented in Paper III and in this paper (see the
following section). Therefore, we treat it as normal outburst
based on its duration and fading rate.

There are apparent oscillations in all days except in January 2.
Figure 4 shows the period-theta diagrams derived from PDM method
(Stellingwerf 1978) for two segments of data, (a) covering
December 30, 31 and January 1 (data A hereafter), (b) covering
January 3 and January 4 (data B hereafter). They give periods,
P$_{A}$ (0.06562(14) d) and P$_{B}$ (0.06638(35) d),
(3.1${\pm}$0.2$){\%}$ and (4.3${\pm}$0.5)${\%}$ larger than the
orbital period of 0.06366 d (Thorstensen and Taylor 1997),
respectively. The mean errors of the periods are the most
pessimistic estimations derived from the method of Fernie (1989).
Fitting sinusoidals to the two segments of data gives two
ephemerises,
\begin{eqnarray}
T_0(max)=HJD2451543.14666+0.06562n \\
T_0(max)=HJD2451547.15245+0.06638n
\end{eqnarray}
where T${_0}$ is the time at phase zero of fit sinusoidals, $n$ is
the cycle number. We also computed period-theta diagram for
individual data series. The results of period determination and
full amplitudes of magnitude variation(${\Delta}$${m}$, derived
from fitting sinusoidal to data in Figure 3a) are also listed in
Table 1. Although the errors for data of each day are rather
large, there is a trend that period decreases with time increasing
within each outburst.

We tried to find whether there is a unique stable period
throughout the whole light curve, i.e., whether the difference
between the periods of data A and data B derived above is due to
intrinsic change or period determination error. The period-theta
diagram for the whole curve is shown in Figure 4c. At the first
glance, there is really a period (P$_{C}$=0.06572 d) which fits
the whole light curve. However, we found that it was not the case.
We tried to add (${n{\pm}r}$)${P_C}$ (${n}$ is an integer, ${r}$
is a real number from 0 to 0.5) to time abscissa of data B. There
was a single dominant peak near P$_{C}$ in derived period-theta
diagram when r is approximately less than 0.2. And there are two
peaks at P${_A}$ and P$_{B}$ when ${r}$ is greater than 0.2. It
means that when two segments data are of different period as in
our case, there is a probability of 40${\%}$ to obtain a single
peak. Because the errors of P${_A}$ and P${_B}$ are the most
pessimistic estimations and the difference of them is larger than
the errors, we believe there is no unique period which fits the
whole light curve. In fact, unstable period is a well-known
property of superhump (Warner 1995). The amplitudes of the
oscillations range from 0.12 mag in December 30 to 0.48 mag in
January 1. Considering the periods and amplitudes of the
modulations, we identify them as superhumps.

Figure 5 shows magnitude variation versus superhump phase
calculated according to equation (2) and equation (3),
respectively. The light curve of January 4 is double-peaked. This
feature suggests that the structure of the accretion disk is more
complex than a simple eccentric disk in the day.

\subsubsection{The normal outburst occurred in 2004}
Figure 3b is the light curve obtained during 2004 February 23-29.
The full amplitude of this outburst is about 2.5 mag. The average
rising and decline rate are 2.5 and 0.7 mag day$^{-1}$,
respectively. It is easily obtained that the period of this normal
outburst is about 6 days, which is 2 days longer than the usual 4
days (Paper I; Paper II; Paper III). From the light curve of
AAVSO\footnotemark\footnotetext{http://www.aavso.org}, we can find
that the mid time of our observation was just about 10 days before
the occurrence time of a superoutburst.

Figure 6 shows the daily light curves with linear trend removed.
The periodgram and window spectrum (Scargle, 1982) of data
covering February 25-28 are shown in Figures 7$a_{1}$ and
7$a_{2}$, respectively. There are two peaks at frequencies 13.938
circles day$^{-1}$ and 14.956 circles day$^{-1}$. But from the
periodgram of data obtained in February 25, as shown in Figure
7$b_{1}$, we can see there is only one high peak at a frequency
14.566 cycles d$^{-1}$. So we believe that the frequency of 14.956
cycles d$^{-1}$ is the true superhump frequency and the frequency
of 13.938 cycles d$^{-1}$ is an alias and the corresponding period
of 0.06686(09) d is (5.0$\pm$0.15)\% larger than the orbital
period.

As shown in Figure 6, the modulations in February 25, when the
observational run was just after the outburst maximum, are the
smallest. The middle panel in Figure 8 (a enlarged view of Figure
6) shows more clearly that the magnitude variation is about 0.05
mag. Moreover, the superhump amplitude is increasing till near the
end of the outburst (see Table 1). The light curve shown in the
lower panel in Figure 8 shows that the superhumps in February 29
are diminishing. At the beginning of the run, there appears to be
a modulation with full amplitude about 0.4 mag. Then it becomes
0.3 mag and till about 0.15 mag at the end of the run. The
periodgram and window spectrum are shown in Figures 9a and 9b,
respectively. The high peak is at a frequency 16.937 circles
day$^{-1}$. It seems to be a negative superhump with a period of
0.05904(55) day, (7.3$\pm$0.9)\% smaller than the orbital period.
Paper III found negative superhump during early rising of a
superoutburst with a period 0.0589 days, which is very consistent
with our result.

\subsection{Fading superhump in the rising to normal maximum}
Figure 5 shows that magnitude variation in January 2 is the
smallest. A magnified view of the variation is shown in Figure 10.
Within the first superhump period, there is a modulation of about
0.06 mag. Within the second superhump period, there is very small
variation approximate to measuring error if ignoring linear trend.
Even considering this trend, the amplitude is much smaller than
that in the first superhump period. Therefore, it is possible that
the superhump begins to appear near the outburst maximum and
disappear before next outburst maximum.

Figure 3b shows the light curve in February 24 is the original
data. At the first glance the modulations only exist at the
beginning of the run. In fact, after cubic polynomial trend
subtraction, as shown in the upper panel in Fig 8 (the magnified
view of Figure 6), the modulations of about 0.1 mag exist along
the whole run. We can see roughly in Figure 8 that the period of
modulation at the second half of this run became about two times
greater than the one at the first half. Period analysis indicates
crudely that these two periods are 0.07516(155) d and 0.11857(334)
d, respectively.

According to the magnitudes and rising rates in paper I and II,
Figure 3b shows rough positions of their observations in a normal
outburst (the symbol ``star" represents estimated position of
maximum light). The observation of paper I mentioned above is
taken in the middle of the rise to the maximum of a normal
outburst, which is very similar to our observation on February 24.
And it can be seen that a modulation (${\sim}$0.1 mag) presents in
the light curve (see 1994 March 11 in their Figure 8). This is
consistent with our light curve shown in Figure 3b. Moreover, a
close inspection of Figure 5 in paper II suggests that the
observing run is also in the rise and right before the maximum of
the normal outburst. A wave (${\sim}$0.05 mag) with a period near
superhump period is gradually disappearing. From Figure 5 of paper
III, we can see that the rise to the normal maximum occurred in
daytime and they missed it. So the observations of the three
groups are all coincident with the observational results in this
paper. Combining the facts described above and this paper, it
seems to be that superhump begins to appear near normal maximum
and disappears before the next normal maximum.

Therefore, superhump behavior described in this paper clarifies
the problem (see $\S 1$). First, it confirms that there are
actually superhumps in normal outbursts of ER UMa. Second,
superhump in the decline part might set out near (maybe at or a
little later than) the normal maximum. Third, superhumps possibly
disappear before the next normal maximum.

\section{Discussion}
In SU UMa-type dwarf novae, two types of humps, which have periods
slightly longer than the orbital periods, have been observed.
These humps are (ordinary) superhumps (Warner 1995) and late
superhumps (e.g., vogt 1983; Hessman et al. 1992). In this
section, we discuss the nature of our observed humps during normal
outbursts in ER UMa.
\subsection{The late superhumps}
In the manner of Vogt's (1982) early model for ordinary
superhumps, Osaki (1985) and Whitehurst (1988) put forward that
the eccentric disk survives for several days after the end of a
superoutburst. Thus, the modulation of the hotspot brightness,
which is caused by the stream impacts the disc at varying depth in
the white dwarf potential, produces late superhumps. But Hessman
et al. (1992) pointed out that the eccentricity determined in OY
Car is too small to produce sufficient modulation. They suggested
that there should be considerable variation in surface density and
scale height around the rim of an eccentric disc, and this should
cause variation in the appearance of the hotspot. However, Rolfe
et al. (2001) find out that brightness of the hotspot varies with
$|\Delta\mathbf{V}|^2$, which supported the former model.

According to the hotspot model, the modulations would diminish in
the rising phase of an outburst and develop in the fading phase.
This is due to that the emission from the accretion disk becomes
more dominant after the onset of outburst. Thus, the contribution
from the hotspot is getting weaker and weaker. In the fading
phase, a reversed effects happen. In this respect, our
observational phenomena could be taken for late superhumps.
However, we cannot make a conclusion that they are late superhumps
since (1) there is no superoutburst observed; (2) the 0.5 phase
shift, which is the basic definition of late superhump, cannot be
investigated.

\subsection{The ordinary superhumps}
The TTI model was proposed to explain bimodal outbursts of SU UMa
stars (Osaki 1989). Many numerical simulations (Hirose \& Osaki
1990; Whitehurst 1994; Kunze et al. 1997; Murray 1998; Truss et
al. 2001) showed that thermal-tidal instability model was very
successful on explaining behaviors of SU UMa stars. Based on the
model and adopting a mass transfer rate 10 times higher than that
expected in the standard theory of the evolution of cataclysmic
variables, Osaki (1995) reproduced the light curve of ER UMa.
However, it should be noted that these simulations only can
demonstrate that the model is able to reproduce the phenomena
known in the past, in which the most important is existence of
superhumps only after supermaximum of SU UMa stars.

Paper III found superhumps in normal outburst and suggested that
superhumps existed essentially in all phase of a supercycle.
Hellier (2001) suggested that eccentricity might last after the
end of superoutburst in low mass ratio systems and the long-life
eccentricity caused superhumps observed after the end of
superoutbursts. Based on the observations in paper III,
Buat-M{\'{e}}nard and Hameury (2002) supported this kind of
``decoupling" and proposed normal superhumps in ER UMa stars
should be ``permanent" caused by permanent eccentricity.

We find that superhump is both in the rise and decline part of a
normal outburst. Our observation also demonstrates that superhumps
disappear in a part of the rise to normal maxium. If the observed
humps were ordinary superhumps, these suggest that superhump
(eccentricity) undergos development and disappearance in a cycle
of normal outburst. In this case, long-life eccentricity can't
explain superhumps in normal outbursts of ER UMa. Superhump is not
permanent in ER UMa, either. Because superhumps develop in each
normal outburst, we have to investigate the relation between
eccentricity (superhump) and superoutburst further.

More observations of other ER UMa-type stars should be taken to
examine whether superhumps in normal outbursts are common in this
kind of system and to examine whether the superhumps really
disappear in a part of the rise to outburst maximum. Second,
further investigations on the relation between superhump and
superoutburst are needed.

\emph{Acknowledgments}:  The authors are grateful to the anonymous
referee for his/her carefully reading of the manuscript and
thoughtful comments. We would like to thank the Optical Astronomy
Laboratory, Chinese Academy of Sciences and Dr. Peisheng Chen of
Yunnan Astronomical Observatory for scheduling the observations.
This work is supported by grants 10173005 and 10221001 from the
National Natural Science Foundation of the People's Republic of
China.

\newpage
\begin{figure}
\centering
\includegraphics[width=13cm]{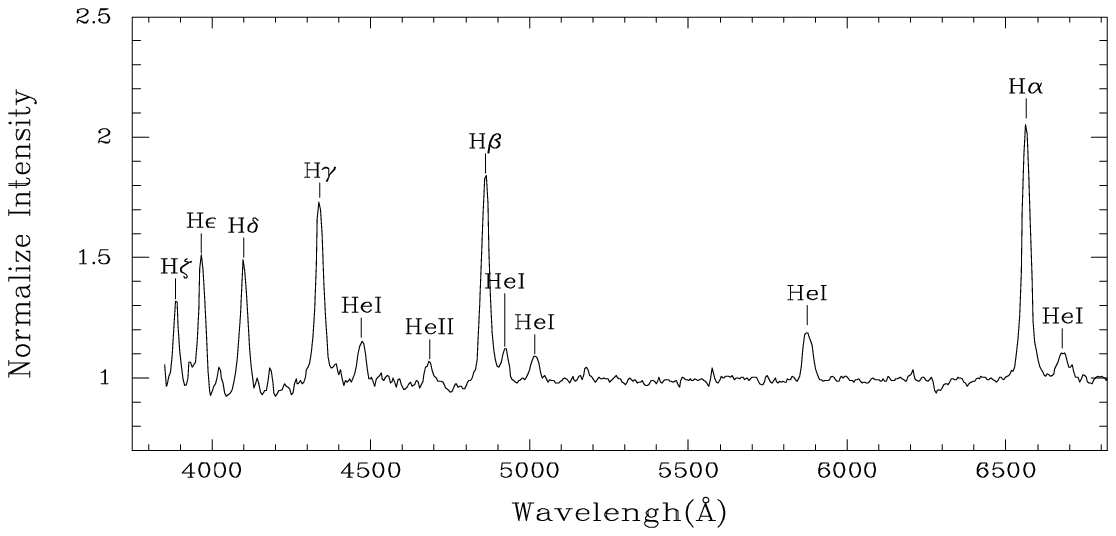}
\caption{The sum of all 63 spectra, showing strong Balmer emission
lines. The helium emission lines from $\lambda$4471 to
$\lambda$6678 can be also seen clearly.}
\end{figure}

\begin{figure}
\centering
\includegraphics[width=13cm]{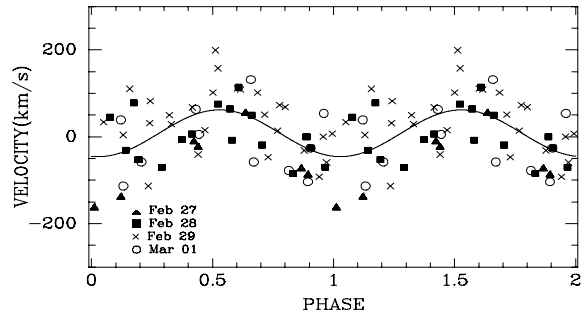} \caption{Radial
velocity of the emission line H$\beta$ and best fit sinusoidal.
Phase computed according to the ephemeris expressed by eq. [1].
The velocities measured in different days presented in the figure
with different symbols, from which we would like to see the
variation of the velocities when the disk at its different
precessing phase. }
\end{figure}

\begin{figure}
\includegraphics[width=13cm]{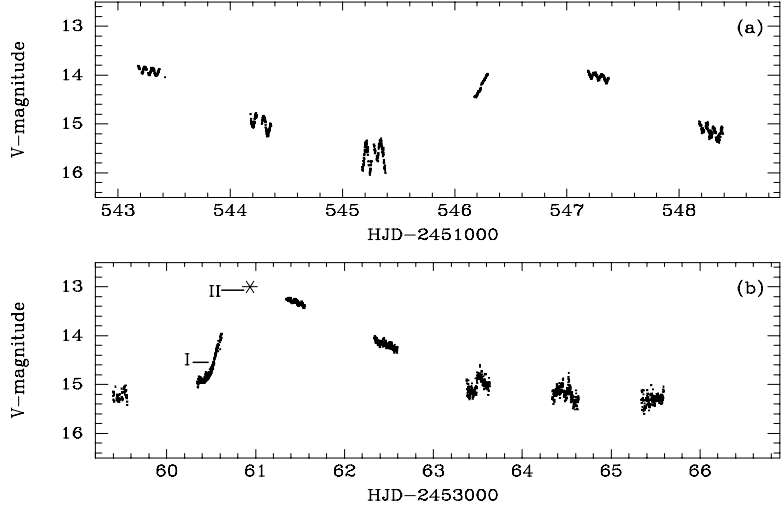} \caption{All of
our photometric data. (a) Data recorded in 1999 December and 2000
January. It shows the decline part of a normal outburst and
another almost complete normal outburst. Modulation of 0.1-0.5 mag
clearly present in the light curve except in January 2. (b) Data
recorded in 2004 February. It shows a complete normal outburst. We
can see clearly that the period of this normal outburst is 6 days,
which is 2 days longer than usual period of ER UMa. The total
amplitude of this outburst is about 2.5 mag. The symbol ``star"
represents estimated position of maximum light. The horizontal
lines, `I' and `II' roughly represent phase of observations of
paper I and paper II in a normal outburst, respectively.}
\end{figure}

\begin{figure}
\includegraphics[width=13cm]{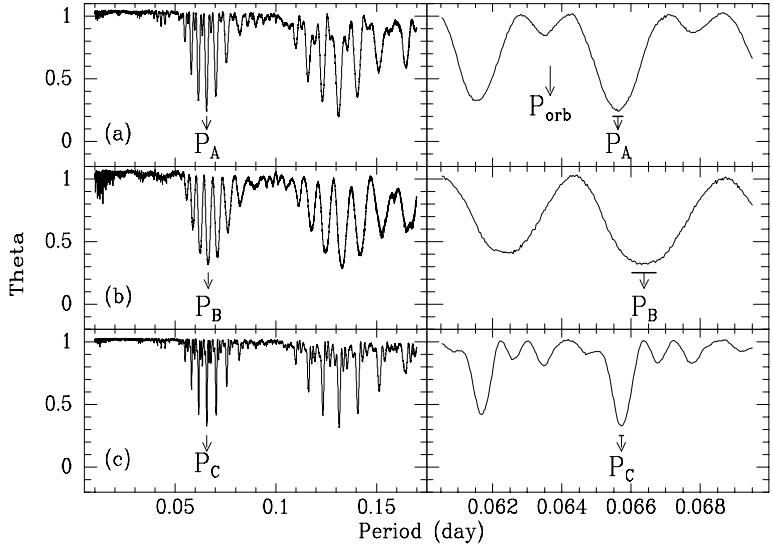}
\caption{Period-theta diagrams of three series of data: (a) From
December 30 to January 1, (b) From January 3 to January 4, (c) the
whole light curve excluding January 2. In the right is magnified
view of each diagram. The horizontal lines at the ends of the
arrows represent the most pessimistic error bars, so we identify
the modulations in the light curve as superhumps. Although there
is a single peak in (c), we believe that there is no unique period
can span the whole light curve (see the text).}
\end{figure}

\begin{figure}
\includegraphics[width=13cm]{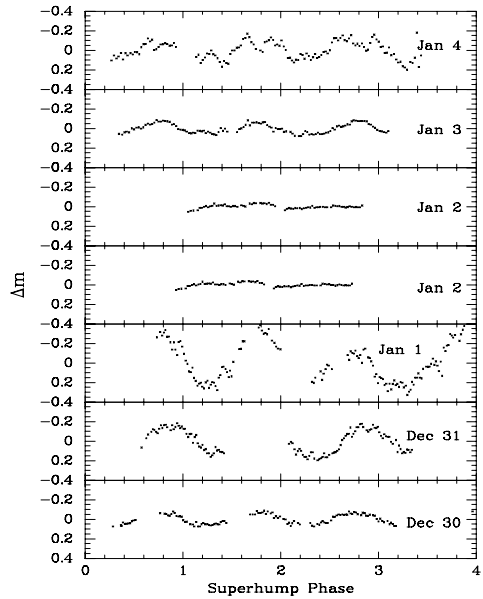} \caption{Magnitude
variation versus superhump phase. Linear trend has been removed.
Superhump phase
 in the lower 4 panels are computed according to equation (2), that
in the higher 3 panels are computed according to equation (3).
Modulation in January 2, if any, is the smallest.}
\end{figure}

\begin{figure}
\includegraphics[width=13cm]{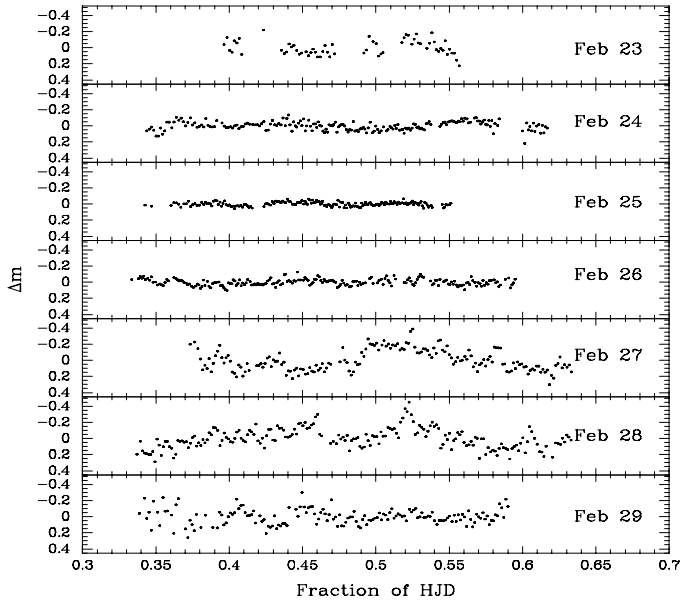} \caption{Daily
light curves recorded in 2004 February. Linear trend has been
removed from all data series except that a cubic polynomial trend
has been subtracted in Feb 24. Modulations of $\sim$0.1 mag and
0.05 mag clearly present in February 24 and 25, the time rising to
and after normal maximum, respectively. And the amplitude
increases during the decline part of the outburst. In February 29,
a wave ($\sim$0.2 mag) is gradually disappearing.}
\end{figure}

\begin{figure}
\includegraphics[width=13cm]{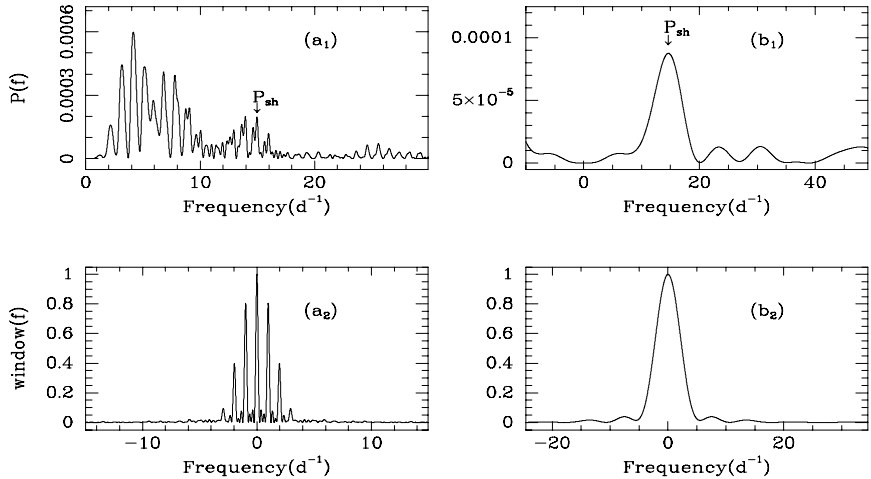}
\caption{Periodgram of data series from February 25 to 28.
($a_{1}$) The power spectrum of data series covered February
25,26,27,28 (data C). ($a_{2}$) The window spectrum of data C.
($b_{1}$) The power spectrum of photometric data recorded in
February 25 (data D). ($b_{1}$) The window spectrum of data D.}
\end{figure}

\begin{figure}
\includegraphics[width=13cm]{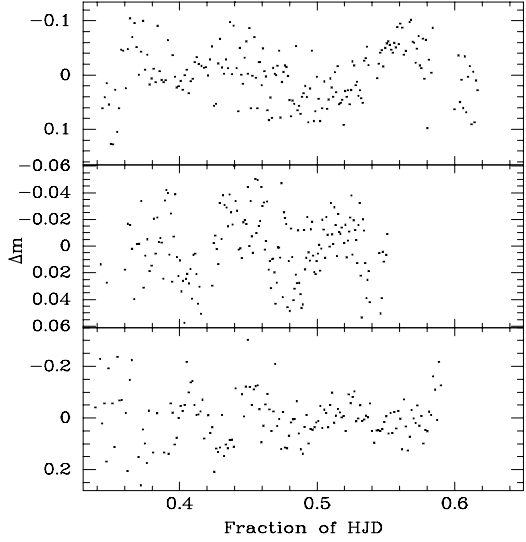} \caption{A
magnified view of magnitude variation in February 24 (upper
panel), February 25 (middle panel) and February 29 (lower panel)
shown in Figure 6. Although modulations present in all three light
curves, the amplitudes are obviously varying. In February 24, the
full amplitude is about 0.1 mag, but it becomes 0.05 mag in
February 25, just before and after the outburst maximum
respectively. In February 29, when the ourburst is close to the
end, modulations are weakening quickly. The full amplitude is from
0.4 mag to 0.15 mag. }
\end{figure}

\begin{figure}
\includegraphics[width=13cm]{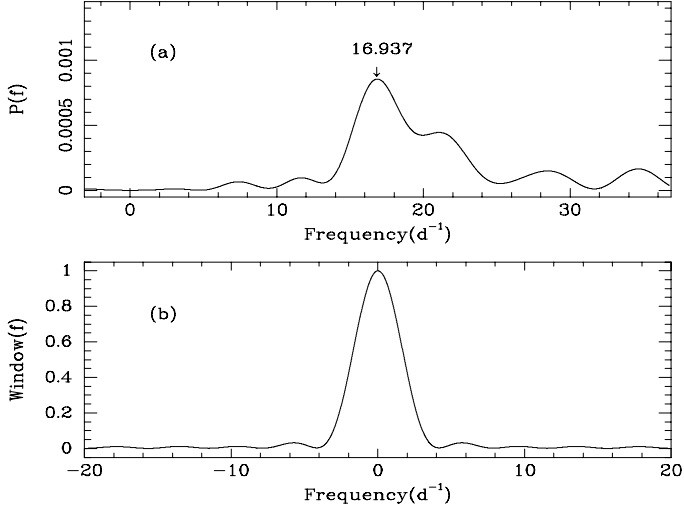}
\caption{Periodgram of photometric data obtained in February 29.
(a) The power spectrum. (b) The window spectrum.}
\end{figure}

\begin{figure}
\includegraphics[width=13cm]{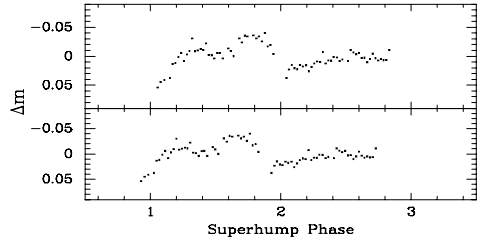} \caption{A
magnified view of magnitude variation in January 2 shown in Figure
5. There is a modulation (${\sim}$ 0.06 mag) in the first
superhump period and no modulation is detected in the second
period.}
\end{figure}
\end{document}